\documentclass[fleqn,twoside]{article}
\usepackage{espcrc2,epsfig}

\usepackage{graphicx}
\usepackage[figuresright]{rotating}

\newcommand{\AmS}{{\protect\the\textfont2
  A\kern-.1667em\lower.5ex\hbox{M}\kern-.125emS}}

\title{Selected design issues of a dedicated facility for generic QCD studies}

\author{{\bf Mieczyslaw Witold Krasny}, L.P.N.H.E, Universities Paris VI et VII,
        4.pl Jussieu, 75252 Paris, France        }
       
\begin{document}

\begin{abstract}  
\vspace{1pc}
QCD, even if presently out of fashion, deserves a  dedicated, generic research program 
providing new challenges for the theory and aiming at understanding hadronic matter  
and vacuum in terms of quark and gluon degrees of freedom. Such a research program needs 
a dedicated facility to re-address basic questions which remain unanswered and to 
open new vistas.    

\end{abstract}

\maketitle

\section{Introduction}

QCD is
the theory of strong interactions. 
At distances, which are sizeably smaller than the size of hadrons, interactions 
of quarks and gluons, the basic degrees of freedom of the theory, 
are precisely controlled by perturbative calculations. At distances
comparable to the size of hadrons, 
the basic degrees of freedom and the symmetries of the theory are hidden
and perturbative methods are no longer valid. Understanding the relationship
between the quark-gluon and the hadronic (nuclear) degrees of freedom 
is the most challenging issue in the domain of strong interactions.
 
The consequences of non-abelian properties of QCD 
can be only partially accessed by the lattice 
calculations. This is both due to the limitation of the power of the
available computers and, more importantly,  due to the very nature of numerical
methods. Further progress in understanding 
the relationship between quarks and gluons and their effective collective 
modes (hadrons and nuclei) is thus bound  to be driven by a generic experimental
program, which needs to develop its novel femto-technology tools,   
allowing not only for observing, but also for 
manipulating quarks and gluons in various QCD media.

Leptons  and nuclei 
are indispensable for development of such tools.
Deciphering  high energy collisions of nucleons and nuclei in terms
of basic quark-gluon processes is difficult.    
Collisions of   nucleons and nuclei with leptons
could play a role of the  Enigma machine capable of  decoding 
the p-A and A-A collision data.
Lepton beams resolve 
interactions of charged constituents of 
hadronic matter with precisely controlled resolution 
power not only at small but  
also at large  (confinement scale)
distances. They can thus be considered as surgery tools
for effective  
filtering of various distance scales involved in the interactions and for
controlling the initial configuration of quark degrees of freedom.
In large coherence-length processes,
nuclei provide effective means to tune 
the projected densities of quark-gluon systems
and/or to tune  the effective strength of colour forces.
Conversely,  
in small coherence-length processes, they play a role 
of variable-length-absorbers of quarks and gluons.
Last but not least, 
nuclei can be used as femto-detectors to observe the space-time
structure of interactions of quarks and gluons. Polarization of 
leptons, protons and light ions
is vital for studies of the spin structure of various QCD media.

Following the discovery of quarks and gluons,
a large number of experiments have been performed to understand their
interactions.  
A future, generic QCD research program at a dedicated
facility, providing collisions of 
large-intensity and atomic-number-tunable ion beams 
with large-intensity polarized beams of protons
and electrons, could extend the frontier of understanding 
in a significant way.
It requires a  research and development effort on 
beam cooling techniques and on high intensity polarized
electron sources.
The progress in collider capacity has to be
matched by a novel detector design  
providing  a full reconstruction
of all particles produced in collisions
of electrons, protons and nuclei.

Such a QCD-dedicated program represented, in my view,
the best  long term research option for the HERA  
collider \cite{HERAeA}, underlying  both the specificity 
and the complementarity of HERA 
with respect to Tevatron and LEP and RHIC.  
This option was considered, however, less attractive than the 
presently realized high-lumi option and it was abandoned.
The eRHIC project \cite{eRHIC} aiming at adding an electron linac
to the existing BNL facilities 
revived and extended the scope of such a program. 
Building the electron linac
at the BNL site have numerous advantages.
For example, a use can be made of   
the available 12 o'clock
collision zone of the RHIC collider 
as a site for the future multipurpose detector.
Such a detector could be optimized not only 
for measuring  collisions 
of protons and nuclei with electron but 
also for measuring collisions of polarized
protons as well as  collisions of protons with nuclei.

The eRHIC design have been 
discussed at several workshops and conferences.
In this short note I sketch  the basic criteria specifying 
the ep (eA) collider parameters  and 
present the first design trial of a dedicated full event detector
for the eRHIC project. 
More details can be found e.g. in talks and lectures on this subject
(the most recent ones can be found in \cite{mytalks}).



\section{Optimizing the ep (eA) collider parameters}

\subsection{Collision Energy range and tunability }

  At high energy frontier, proton-proton and proton-nucleus 
  collisions provide already an adequate environment to study  
  short-distance interactions of quarks and gluons. 
  The low-x, large coherence-length  spin-
  and momentum-structure of nucleons 
  can be resolved 
  using intermediate 
  energy electrons. For example, a 10 GeV 
  electron beam colliding with a 100 GeV protons 
  (nuclei) resolves 
  the deep inelastic structure of the target 
  down to $x_{Bj}=10^{-3}$.
  At this $x_{Bj}$ value the   
  coherence-length exceeds already by one order of magnitude the  
  size of heaviest nuclei allowing for a sufficient lever
  arm for extrapolations to even smaller $x_{Bj}$  values.

          The ep (eA) centre-of-mass collision energy range optimal for  
          covering  both the perturbative QCD region and 
          the distance scales relevant for hadronic
          matter such as the proton radius - $R_p$, nucleus radius - $R_A$, 
          $1/ \Lambda_{QCD}$, 
          $1/m_{\rho}$ is $\sqrt{s} = 10-100$ GeV.
          The above energy span  provides an  
          optimal  resolution range of the electron probe
          both in the transverse and in the collision
          (light-cone) direction. The low energy limit can be
          approached in parasitic fixed target collisions 
          of the electron beam and thus can be moved above 10 GeV. 
          Tunability of the 
          beam collision energy 
          is indispensable to filter out 
          processes mediated by the transverse and the longitudinal
          virtual photons and is vital in understanding 
          quark and gluon energy loss in various QCD media.       

\subsection{Ratio of lepton and nucleon/ion beam energies}

          Choosing the ratio
          of the proton (ion) to electron beam energies 
          in the range 10-20 facilitates, in particular 
          at the highest centre-of-mass energy,  various aspects
          of the full event detector design. It  assures both 
          a good reconstruction quality and an efficient triggering of
          deep inelastic scattering events over the full $x_{Bj}$
          range.  
          In addition, such 
          a value of 
          the beam energy ratio allows for an early decoupling 
          the electron and of the  proton (ion) beam optics, thus simplifying 
          the design of the electron beam insertion.
          Last but not least, it permits operating high electron 
          currents with controllable level of synchrotron radiation
          at the interaction point. 
           
\subsection{Charge and polarization of the electron beam}

          While highest and precisely controllable polarization of
          the electron and of the proton (light ion) beam is indispensable,
          a possibility of switching from electron to positron beam is 
          of secondary importance. A majority of QCD processes
          leading to the charge asymmetries can be investigated
          by looking at spin asymmetries.

\subsection{Choice of ion species} 

          It is mandatory to collide electrons with  heavy ions 
          in order to maximize 
          the nuclear medium effects and 
          with lightest ions (e.g. 
          deuteron) in order to  study proton/neutron flavour asymmetries.
          In addition, it is highly desirable to cover uniformly the 
          A-range on the $A^{1/3}$ scale. Isoscalar ions 
          should be preferably chosen, especially if the 
          collider compaction factor will  permit  simultaneous
          storing of two, isoscalar ion bunch-trains.

\subsection{Luminosity}

          While highest achievable luminosity $L*A \approx 10^{33}~ [cm^{-2}s^{-1}]$ is necessary 
          for high precision measurements of spin asymmetries,
          for studies of exclusive and semi-exclusive processes, for
          rare fluctuations of nuclear densities and  for the $x>1$
          physics, there exist  a vast inclusive physics program,
          which can be executed already at the luminosity, which is three orders
          of magnitude smaller than that quoted above.
          This creates an opportunity to consider the collider/detector project
          as evolutive, containing several open scenarios. 
          Such a flexibility allows, for instance, 
          for running at lower luminosities at comfortable $\beta^{*}$
          to study  the electron, proton and ion fragmentation
          processes.

\subsection{Beam parameters at fixed luminosity}
          
          It is desirable to 
          minimize beam emittance rather than $\beta^{*}$
          at the Laslet and at the beam-beam tune 
          shift luminosity limits.
          As a thumb rule the beam divergences at the interaction point
          should be kept largely below the level of 
          $P_{Fermi}/P_{Beam} = 5 *  10^{-4}$
          for unambiguous identification of interacting nucleons
          with respect to spectator ones.     
          In order to assure  efficient triggering over the full $x_{Bj}$-range   
          an emphasis
          has be put on cleanness of the beams, i.e. on the smallest possible
          halo of ``out-of-bunch'' particles.
 
\subsection{Interaction Point geometry}

          Merits of the proposed range of collision energies, 
          of the beam energy ratio and of using electron
          linac rather than circular machine are clearly visible 
          while designing the  
          interaction point geometry.
          In the configuration discussed above a clash  between the
          magnetic lattice of the  
          electron beam insertion and that  
          of the proton (ion) beam can be largely avoided.
          In the limit of small ratio
          of electron to proton (ion) beam energies
          the electron insertion magnet causes only  residual
          distortion of the proton (ion) beam trajectories.
          At small electron beam energy,    
          it can  be placed in the vicinity of the beam 
          crossing zone and used 
          as a momentum analyzer of particles
          produced at small angles while 
          avoiding severe constraints related  to 
          effective masking of synchrotron radiation.
          The layout of the interaction point geometry 
          is greatly simplified if pre-polarized      
          electrons are accelerated in a linac
          with no need for a spin rotation.

 \begin{figure}[!htb]
\vspace{1pc}
 \centering
 \epsfig{file=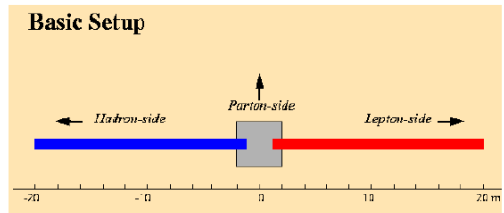,bbllx=123pt,bblly=323pt,bburx=487pt,bbury=466pt,
         width=4.3cm,angle=270}
\label{setup}
\caption{The detector geometry.}
\end{figure}

\subsection{Three beams collider}

          The RHIC collider provides
          beams of ions and polarized protons, which cover
          fully the optimal energy range
          for the future ep (eA) collider (lower energy ion 
          and proton beams  can be stored
          at RHIC if cooled by a low energy  
          electron beam).
          Thus, the  Brookhaven National Laboratory 
          turns out to be a cost effective site to conduct 
          the future ep (eA) research program in close synergy with the present 
          RHIC pp, AA and pA programs.       
          An ambitious but extremely interesting scenario
          is to design the IP12 collision point as a collision point of 3 beams.
          In such a scenario one of the ion (proton) beam or the electron beam is guided 
          in and out of the beam-beam collision zone, within a common 
          beam tube, allowing for measuring ep, eA, pp, and pA 
          collisions in the same region in  a common full event detector.
          The ep (eA) collisions could thus be observed with no 
          interference with the ongoing RHIC program. 
          More importantly, a precise relative studies of 
          medium-dependence of basic QCD processes can be made 
          by using the same detector for ep, eA, pp and pA collisions.
          The collider parameters discussed above, in particular 
          the particular the choice of the range of electron energies with respect
          to the energies of RHIC beams facilitate such a design.

\section{Full event detector - selected design issues}

The full event detector has to  provide measurements of all 
particles produced in the collisions.
Compared to 
the HERA or to the Tevatron detectors, it extends the measurement region 
covering the soft remnants
of electrons, nucleons and nuclei. 
It redefines the conventional  detector-machine interface.

\begin{figure}[!htb]
\vspace{1pc}
 \centering
 \epsfig{file=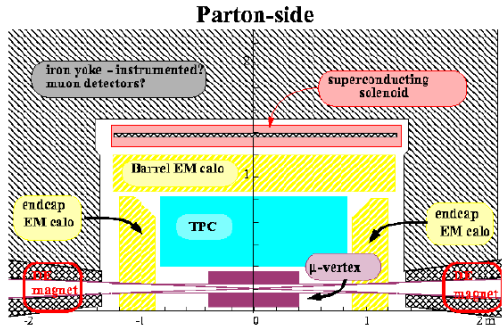,bbllx=123pt,bblly=323pt,bburx=487pt,bbury=466pt,
         width=4.3cm,angle=270}
\label{parton}
\caption{The parton side of the generic detector}
\end{figure}

The first attempt to design 
such a detector has been 
presented at the Yale workshop \cite{krasnydet}.
The following criteria were imposed 
for this initial trial: 
\begin{itemize}

\item 
 The detector should be common 
  for ep, eA, pp and pA collisions
  to study both  the  
  hard scattering
  of partons and  the beam particle dependent medium effects.
\item 
It should allow for the 
 reconstruction of complete ep and eA events 
 (covering the proton fragmentation region in the pp and the pA collisions,
 a provision for adding a
 muon detection system and a charged kaon identification system, 
 remains open at this design stage).
\item
The beam crossing optics should  minimize 
the clash 
with the existing RHIC interaction region optics.
\item  
The existing RHIC magnets 
should  be used, as much as they can be useful, 
  in the spectrometer design.
\item 
The electron insertion should provide a  functionality 
of the  spin rotator. 
\item 
The design should remain invariant of  
the choice of Ring-Ring versus  Linac-Ring collisions.
\item 
The detector should provide 
precise luminosity monitoring and 
good experimental control of radiative corrections 
for electron induced reactions.
\end{itemize}

 \begin{figure}[!htb]
 \centering
 \epsfig{file=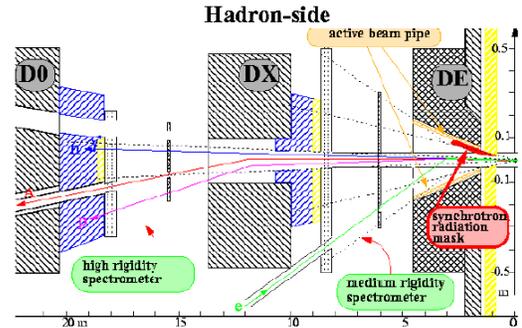,bbllx=123pt,bblly=323pt,bburx=487pt,bbury=466pt,
         width=4.3cm,angle=270}
\label{hadron}
\caption{The hadron side of the generic detector}
\end{figure}

The general layout of the detector geometry is shown in Fig. 1.
It is symmetric around the collision axis with a central detector
and two symmetric arms along the z direction. For the ep and eA
collisions the instrumentation of the left arm, called hereafter
the hadron side, and the right arm, called hereafter the electron
side is initially asymmetric. It would be possible to 
upgrade such a detector for pp and pA collisions by adding 
the detector elements of the hadron side to the electron side.

The function of the central detector, shown in Fig. 2,
is to determine the kinematics of hard processes 
by measuring the momenta and angles of outgoing leptons
quark and gluon jets. The barrel is a TPC backed by a gas EM 
calorimeter inside a super-conducting coil. Both end-cups are
SPACAL calorimeters. This is a minimal set up for the 
parton detector.

The proton (nucleus) fragmentation region is covered  by the 
high rigidity and medium rigidity spectrometers
shown in Fig. 3. The
function of these 
spectrometers is to measure wounded and evaporated nucleons, nuclear 
fragments as well as other low-angle particles. The spectrometers 
involve the tracking modules and calorimeter stations and
use, in measuring the momentum of charged particles, the 
magnetic field of the D0 and DE magnets.

 \begin{figure}[!htb]
\vspace{1pc}
 \centering
 \epsfig{file=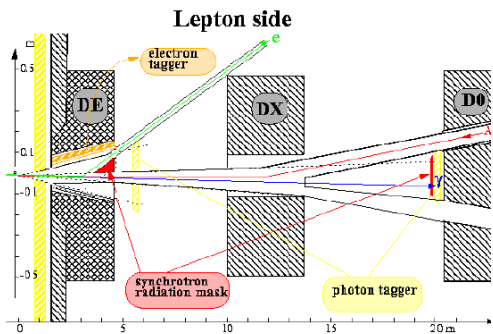,bbllx=123pt,bblly=323pt,bburx=487pt,bbury=466pt,
         width=4.3cm,angle=270}
\label{lepton}
\caption{The lepton  side of the generic detector}
\end{figure}

The detection system in the electron fragmentation region, 
shown in Fig. 4, defines the minimal setup for the 
electron-nucleon (electron-nucleus) collisions. 
It provides a  precise tagging of DIS and photo-production 
processes and measures the low angle photons for luminosity 
measurement and provides experimental means to control the
electro-magnetic radiative corrections.

The spectrometer optics combines several functions.
The 2.3 Tm bending power of the DE magnet distributed over 3 meters
is 3 times higher than the inflection field at DESY. It produces
a deflection angle of 70 mrad for 10 GeV beam and is
designed to rotate the spin vector by 90 degrees (this
constraint is only of importance for a circular electron
machine). With this geometry the electron beam bypasses the DX
magnet entirely, thus uncoupling the electron beam optics
from the ion beam optics. The geometry of the present design have,
unquestionable advantages with respect to that of the HERA 
collider. Owing to smaller electron beam energy the eRHIC 
interaction region can tolerate 10 times higher electron beam current
than HERA i.e. 600 mA. Using low emittance electron-linac beam 
a 100 fold increase of luminosity can be achieved while
keeping the synchrotron radiation at tolerable level.


\begin{thebibliography}{9}
\bibitem{HERAeA} M. W. Krasny, Summary talk at the ``Future Physics 
                               at HERA'' workshop, Hamburg 1996,
                               http:// www.desy.de/heraws96. \\
                 M. Arneodo, A. Bialas, M.W. Krasny, T.Sloan, M. Strikmann,
                               in: Proceedings of the``Future Physics 
                               at HERA'' workshop, Hamburg 1996.\\
                 M.W. Krasny,  Electron-Nucleus Collisions at HERA,
                               in: GSI-REPORT 97-04, April 1997.  
         
\bibitem{eRHIC} R. Holt et al., The Electron-Ion-Collider, White Paper
                               submitted to the NSAC Long Range Planning Meeting,
                               BNL, March 2001.
                                
                
\bibitem{mytalks} M.W. Krasny, {\it Physics with Lepton-Ion colliders}, Snowmass 2001,
                               http://www-lpnhep.in2p3.fr/$\sim$krasny/snowmass2001-physics.ps. \\
                               {\it Detector Issues of Lepton-Hadron Colliders}, Snowmass 2001,
                               http://www-lpnhep.in2p3.fr/$\sim$ krasny/snowmass2001-detector.ps. \\
                               {\it The relationship of the lepton-nucleon and lepton-nucleus studies
                               to heavy ion collisions}, Newport 2001,
                               http://www-lpnhep.in2p3.fr/$\sim$krasny/gordon2001.ps. 
                               
               
\bibitem{krasnydet} M.W. Krasny, {\it Physics of eA at eRHIC}, Yale 2000,
                               http://www-lpnhep.in2p3.fr/$\sim$krasny/yale2000.ps.
               
\end{thebibliography}
\end{document}